\documentclass{article}

\usepackage{PRIMEarxiv}

\usepackage[utf8]{inputenc} % allow utf-8 input
\usepackage[T1]{fontenc}    % use 8-bit T1 fonts
\usepackage{hyperref}       % hyperlinks
\usepackage{url}            % simple URL typesetting
\usepackage{booktabs}       % professional-quality tables
\usepackage{amsfonts}       % blackboard math symbols
\usepackage{nicefrac}       % compact symbols for 1/2, etc.
\usepackage{microtype}      % microtypography
\usepackage{lipsum}
\usepackage{fancyhdr}       % header
\usepackage{graphicx}       % graphics
\graphicspath{{media/}}     % organize your images and other figures under media/ folder
\usepackage{amsmath}
\usepackage{amssymb}  % 可选
\usepackage{float}
\title{Non-Hermitian topological electric circuits with projective symmetry
}

\author{
  Wenjie Zhang$^{1}$, Yuting Yang$^{1,2,*}$, Xiaopeng Shen$^{1}$, Liwei Shi$^{1}$,   Zhi Hong Hang$^{3,4}$\\
  $^{1}$School of Materials and Physics, China University of Mining and Technology 
  Xuzhou 221116, China \\
  $^{2}$  State Key Laboratory of Millimeter Waves, Southeast University, Nanjing 210096, China\\
   $^{3}$  School of Physical Science and Technology \& Collaborative Innovation Center of Suzhou Nano Science and Technology, \\Soochow University, Suzhou 215006, China\\
    $^{4}$  Institute for Advanced Study, Soochow University, Suzhou 215006, China\\
    $^*$ Corresponding author:  \texttt{yangyt@cumt.edu.cn} 
}

\begin{document}
\maketitle

\begin{abstract}
Non-Hermitian topological insulators have attracted considerable attention due to their distinctive energy band characteristics and promising applications. Here, we systematically investigate non-Hermitian Möbius insulators and graphene-like topological semimetals from the projected symmetry and realize their corresponding topological phenomena in an electric circuit-based framework. By introducing a nonreciprocal hopping term consisting of negative impedance converters into a two-dimensional electric circuit, we establish an experimental platform that effectively demonstrates that introducing non-Hermitian terms significantly enhances the energy localization of topological edge states, which originate from the non-Hermitian skin effect. Furthermore, a thorough comparison of experimental measurements with numerical simulations validates the robustness and reliability of our electric circuit structure. This work not only reveals the physical properties of non-Hermitian topological materials but also provides valuable theoretical and experimental guidance for the implementation of topological circuits and the design of radiofrequency devices in the future.
\end{abstract}

% keywords can be removed
% \keywords{First keyword \and Second keyword \and More}

\section{Introduction}\label{sec:1}

The symmetry-protected topological phases have been a hot topic in condensed matter physics and artificial periodic systems, such as acoustic and photonic crystals, and electric circuits~\cite{1,2,3,4,5,6,7,8,9}. The topological insulators have gathered significant attention due to their topologically protected edge states, which exhibit remarkable robustness against external disturbances and impurities. Various symmetries, such as the time-reversal symmetry and the point group symmetry, dictate the topological classification and topological band characterization. The recent studies have found that crystal symmetries are projectively represented under a $\mathbb{Z}_{2}$ gauge field (0 and $\pi$ flux over the lattice), which impacts the algebra of spatial symmetry operations~\cite{10,11,12}. The artificial periodic systems with the time-reversal symmetry have intrinsic synthetic gauge field, which means the hopping amplitudes are real numbers that can take either positive or negative signs~\cite{13,14}. The synthetic gauge field enriched the classification of topological phases. The broken primitive translational symmetries give rise to novel M{\"o}bius topological insulators with twisted edge bands and graphene-like semimetal phases with flat bands, which have been realized in acoustic crystals, photonic microring lattices, photonic waveguides and electric circuits~\cite{15,16,17,18,19,20}. 

In a genuine open physical system, non-negligible interactions with the environment lead to a non-Hermitian Hamiltonian, including complex eigenvalues. The interplay between non-Hermitian and topological phases has induced many intriguing phenomena. A recent discovery is the breakdown of the conventional bulk-boundary correspondence, derived from non-Hermitian skin effects, where all eigenstates are localized near the edge of an open system~\cite{21,22,23,24,25}. Non-Hermitian skin effects have been investigated in meta-crystals and quantum walks~\cite{26,27,28,29,30,31}. A non-Hermitian term can be conveniently introduced into an electric circuit~\cite{32,33,34,35,36}, for example, by utilizing a negative-impendence converter with unidirectional current flow to create nonreciprocal hoppings~\cite{37,38,39}. With the help of the excellent boundary sensitivity of non-Hermitian skin effects and flexible construction of electric circuits, it holds the significant potential for designing advanced topological radiofrequency devices, such as a sensor with high sensitivity~\cite{40,41,42}.

Topological insulators with projected symmetry have been extensively investigated in many studies; however, the experimental demonstration for their non-Hermitian properties have not been performed. In this work, we construct non-Hermitian M{\"o}bius topological insulators and graphene-like semimetal phases within a two-dimensional electrical circuit. Nonreciprocal hopping is achieved through the use of negative-impedance converters. Both simulations and experimental results confirm the presence of the non-Hermitian skin effect.

\section{Non-Hermitian Möbius insulator}\label{sec:2}

\begin{figure}[!ht]
    \centering
    \includegraphics[width=0.7\linewidth]{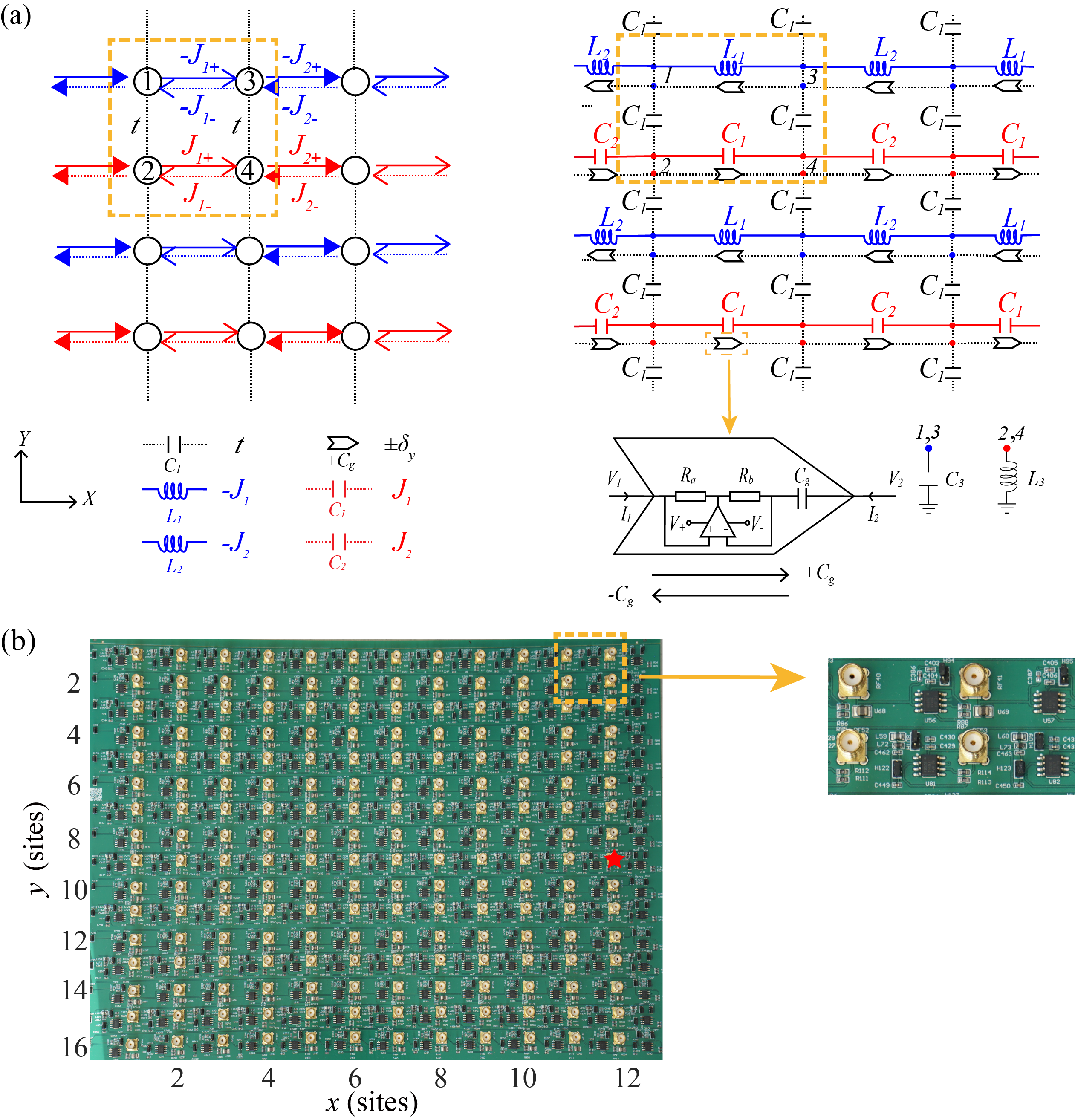}
    \caption{(a) Left panel: Schematic of the tight-binding model of the non-Hermitian Möbius insulator, realized by adding nonreciprocal hoppings. Right panel: Electric circuit realization of the non-Hermitian Möbius insulator through introducing negative-impedance converters. A unit cell indicated by a rectangular box includes four nodes labeled with the numbers. (b) Photograph of the electric circuit board in the experimental measurement.}
    \label{fig:1}
\end{figure}

We firstly consider a two-dimensional rectangular lattice model with $\pi$ gauge flux per plaquette, proposed in Ref.~\cite{11,13}. When the primitive translational symmetries along the $x$- and $y$- directions are preserved, fourfold degenerated Dirac points emerge at the corner of the first Brillouin zone. Breaking the primitive translational symmetry only along $y$-direction through staggered dimerization, the degenerated Dirac points are lifted, resulting in a band gap. The detailed projective symmetry is shown in Supplementary Material. This leads to the proposal of the Möbius insulator, characterized by a twisted edge band dispersion analogous to a Möbius strip. As illustrated in Fig.~\ref{fig:1}(a), the non-Hermitian Möbius insulator is constructed by introducing nonreciprocal hoppings. The parameter $t$ represents the hopping along the $y$ direction, while $J_{1}$ and $J_{2}$ denote the intra- and inter- coupling coefficients, respectively. The nonreciprocal and dimerized hoppings are $J_{1\pm}=J_{1}\pm\delta_{x}$ and $J_{2\pm}=J_{2}\pm\delta_{x}$ along the $x$ direction. The Hamiltonian of this tight-binding model is~\cite{43}:

\begin{equation}
H=\left[ \begin{matrix}
   0 & T_{y} & V+\Delta_{x} & 0  \\
   T_{y} & 0 & 0 & -(V+\Delta_{x})  \\
   V^{*}+\Delta_{x} & 0 & 0 & T_{y}  \\
   0 & -(V^{*}+\Delta_{x}) & T_{y} & 0  \\
\end{matrix} \right]
\label{eq:1}
\end{equation}

where $T_{y}=2t\cos (k_{y}/2)$, $V=-J_{1}e^{ik_{x}/2}-J_{2}e^{-ik_{x}/2}$, and $\Delta_{x}=-2i\delta_{x}\sin (k_{x}/2)$.

The electric circuit realization of the non-Hermitian Möbius insulator is shown in the right panel in Fig.~\ref{fig:1}(a). The hopping parameter $t$ corresponds to the capacitance value $C_1$ between two nodes in the designed circuit configuration. The positive hoppings $J_1$ and $J_2$ correspond to capacitance values $C_1$ and $C_2$, respectively, while the negative hoppings correspond to inductance values $L_1$ and $L_2$. The nonreciprocal term $\delta_{x}$ is achieved via a negative impedance converter with a current inversion (INIC), which incorporates an operational amplifier along with a resistor (denoted as $R_a$) and a capacitor (denoted as $C_2$). This configuration is characterized by a positive capacitance $C_g$ in one direction and a negative capacitance $-C_g$ in the reverse direction. We can obtain the Laplace form of its nodal voltage equation:

\begin{equation}
\left( \begin{matrix}
   I_{1}  \\
   I_{2}  \\
\end{matrix} \right)=i\omega C_{g}\left( \begin{matrix}
   -1 & 1  \\
   -1 & 1  \\
\end{matrix} \right)\left( \begin{matrix}
   V_{1}  \\
   V_{2}  \\
\end{matrix} \right)
\label{eq:2}
\end{equation}

The parameters in the lattice model are set as $t_{1}=J_{1}=1$ and $J_{2}=3.3$. When the hopping ratio $J_{2}/J_{1}=C_{2}/C_{1}=L_{1}/L_{2}$ is greater than 1, the system is a topological Möbius insulator~\cite{15}. Consequently, the values of circuit components are selected as $C_{1}=10\,\text{nF}$, $C_{2}=33\,\text{nF}$, $L_{1}=10\,\mu \text{H}$, and $L_{2}=3.3\,\mu \text{H}$. The inductance values $L_2$ should be 3.03 $\mu$H. To simplify the selection of circuit components for the experimental setup, $L_2$ is chosen to be 3.3 $\mu$H. Owing to the robustness of the topological insulator, the slight change does not affect the circuit characteristics. Nodes 1 and 3 should be connected in parallel to a ground capacitor $C_3$ (43 nF), while nodes 2 and 4 are connected in parallel to a ground inductor $L_3$ (2.5 $\mu$H). The non-Hermitian parameter $\delta$ is defined as the ratio of the capacitance $C_g$ to $C_1$, i.e., $\delta =C_{g}/C_{1}$, where the capacitance $C_g$ varies under the fixed $C_1$. This ratio characterizes the nonreciprocal hopping strength in the system and plays a critical role in the manifestation of the non-Hermitian skin effect.

According to the Kirchhoff current law and Bloch theorem, we can obtain the circuit conductance matrix (detailed derivation in Sec.~\ref{sec:1} within Supplementary Material):

\begin{equation}
J_{m}=i\omega C_{m}+\frac{L_{m}}{i\omega }
\label{eq:3}
\end{equation}

where

\begin{equation}
C_{m}=\left[ \begin{matrix}
   t_{1} & A_{y} & \sigma  & 0  \\
   A_{y} & t_{2} & 0 & B_{x}-\sigma   \\
   \sigma  & 0 & t_{1} & A_{y}  \\
   0 & B_{x}^{*}-\sigma  & A_{y} & t_{2}  \\
\end{matrix} \right]
\label{eq:4}
\end{equation}

with $t_{1}=2C_{1}+C_{3}$, $t_{2}=3C_{1}+C_{2}$, $A_{y}=-2C_{1}\cos (k_{y}/2)$, $B_{x}=-C_{1}e^{ik_{x}/2}-C_{2}e^{-ik_{x}/2}$, and $\sigma =2iC_{g}\sin (k_{x}/2)$,

\begin{equation}
L_{m}=\left( \begin{matrix}
   \frac{1}{L_{1}}+\frac{1}{L_{2}} & 0 & -\frac{e^{ik_{x}/2}}{L_{1}}-\frac{e^{-ik_{x}/2}}{L_{2}} & 0  \\
   0 & \frac{1}{L_{3}} & 0 & 0  \\
   -\frac{e^{-ik_{x}/2}}{L_{1}}-\frac{e^{ik_{x}/2}}{L_{2}} & 0 & \frac{1}{L_{1}}+\frac{1}{L_{2}} & 0  \\
   0 & 0 & 0 & \frac{1}{L_{3}}  \\
\end{matrix} \right)
\label{eq:5}
\end{equation}

To solve for the eigenfrequency of the circuit system, we can let $I_{n}=0$, and get $JV=0$. $I$ is the current and $V$ is the node voltage. We can obtain the eigenfrequency of the non-Hermitian circuit system by solving $\det [J]=0$. This method has some limitations. When the derivative matrix is complex, it is difficult to find the eigenfrequency. By taking the derivative of $V$ for $JV = 0$ with respect to time, we can get the following two differential equations~\cite{34}:

\begin{equation}
-i\frac{d}{dt}\dot{V}=i\frac{L}{C}V
\label{eq:6}
\end{equation}

\begin{equation}
-i\frac{d}{dt}V=-iE\dot{V}
\label{eq:7}
\end{equation}

where $E$ is the unit matrix. The first-order derivative in the time domain is equivalent to multiplying by $i\omega$ in the frequency domain. Hence, we can derive the relationship and obtain:

\begin{equation}
-i\cdot i\omega \left[ \begin{matrix}
   \dot{V}  \\
   V  \\
\end{matrix} \right]=i\left[ \begin{matrix}
   0 & \frac{L}{C}  \\
   -E & 0  \\
\end{matrix} \right]\left[ \begin{matrix}
   \dot{V}  \\
   V  \\
\end{matrix} \right]
\label{eq:8}
\end{equation}

This formula exhibits a formal similarity to the Schrödinger equation, allowing us to make the Hamiltonian matrix equal to:

\begin{equation}
H=\left[ \begin{matrix}
   0 & \frac{L}{C}  \\
   -E & 0  \\
\end{matrix} \right]
\label{eq:9}
\end{equation}

The eigenvalues of the $H(k)$ matrix directly correspond to the angular frequencies of the system. The eigenvalues obtained through the solution of $\det(H)$ are divided into two distinct parts $(\omega ,\omega^{*})$. The real part of $\omega$ is positive and can be utilized as the eigenfrequency of the circuit.

\begin{figure}[!ht]
    \centering
    \includegraphics[width=0.7\linewidth]{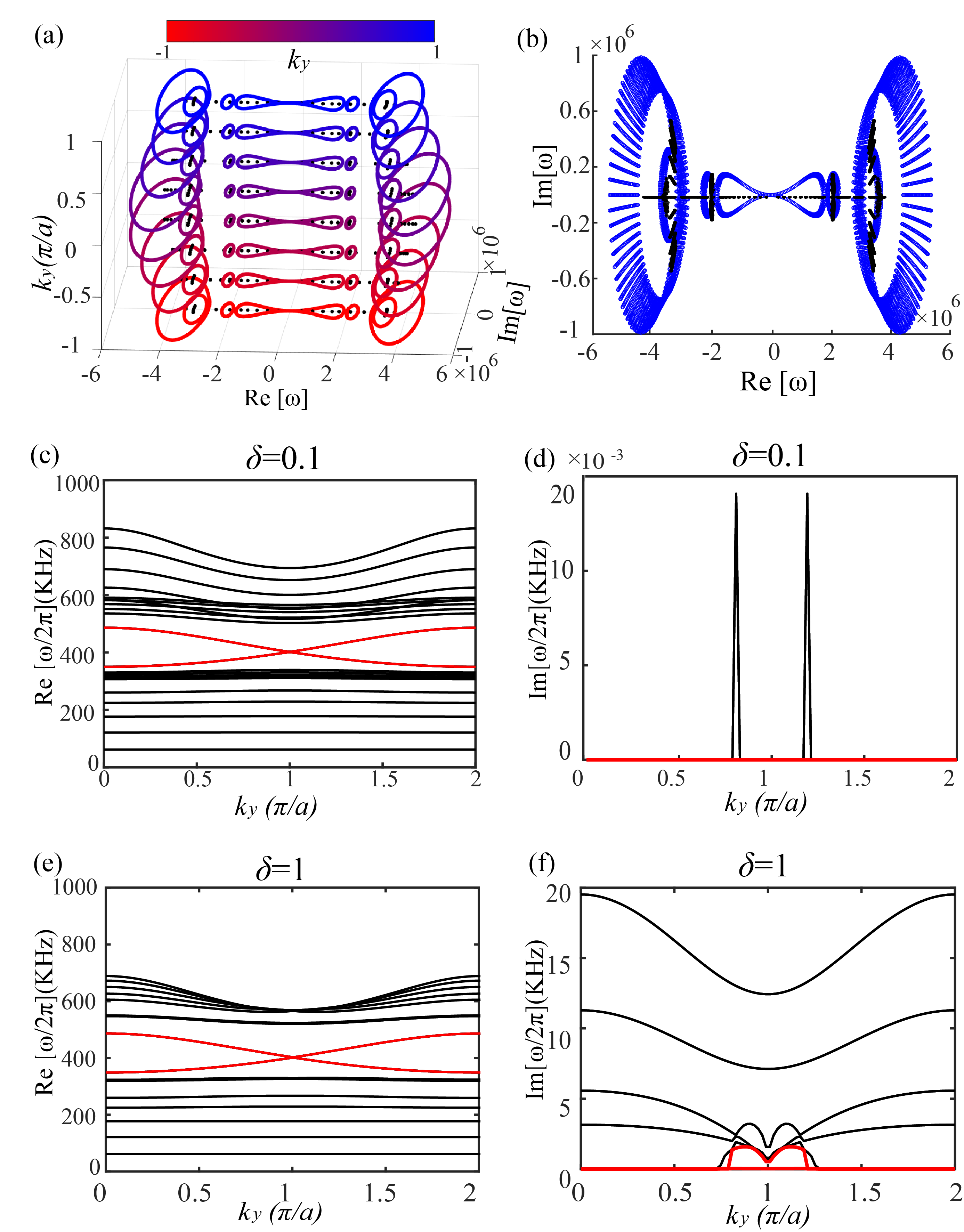}
    \caption{(a) Complex frequency spectra of the non-Hermitian Möbius insulator at different wave vector $k_y$ for the nonreciprocal hopping strength $\delta=1$. The frequency spectra with closed loops correspond to the electric circuit system under x-PBC and y-PBC, while the open arcs are plotted under x-OBC and y-PBC. (b) Projection of complex frequency spectra from panel (a). (c) and (e) Real part of eigenfrequencies with wave vector $k_y$ for $\delta=0.1$ and $\delta=1$ respectively, while considering x-OBC and y-PBC within the non-Hermitian circuit system. The black curves represent the bulk bands, and the red curves indicate the twisted edge bands. (d) Imaginary part of eigenfrequencies for $\delta=0.1$ and $\delta=1$, respectively.}
    \label{fig:2}
\end{figure}

The non-Hermitian skin effect is induced through tuning the nonreciprocal terms in the electric circuit. By calculating the eigenvalues, the theoretical dispersion curves can be obtained under the periodic boundary condition (PBC) and open boundary condition (OBC). The eigenfrequencies under the PBC along x and y directions (denoted as x/y-PBC) form closed loops in the complex plane for each fixed $k_y$, as shown in Fig.~\ref{fig:2}(a). The eigenvalues are plotted under x-OBC and y-PBC, indicated by black lines, which form open arcs corresponding to the non-Hermitian skin modes. The real eigenfrequency of the dispersion under x-OBC and y-PBC are displayed in Figs.~\ref{fig:2}(c) and~\ref{fig:2}(e), corresponding to the nonreciprocal hopping strength $\delta=0.1$ and $1$, respectively. Two twisted edge bands denoted by red lines emerge within the band gap, fully detached from the bulk bands. These real edge bands are similar with those in Hermitian Möbius insulator. When $\delta=0.1$, the non-Hermitian skin effect is relatively weak, and the imaginary part of the eigenfrequency is very small. Under $\delta=1$, the imaginary part of the frequency is large. The ratio $\delta$ varies as a function of frequencies (see Fig.~S3 in Supplementary Material). The nonreciprocal term has a more pronounced impact on the system, leading to a notably enhanced non-Hermitian skin effect.

Subsequently, we construct a matrix consisting of eigenvectors, where each column contains information about all nodes at a specific frequency. We clarify that the eigenvector of the effective Hamiltonian is denoted as 
$
\psi =\left[ \begin{matrix}
   V  \\
   \dot{V}  \\
\end{matrix} \right],
$
and we define 
$
\varphi_{i}=|\psi_{i}|^{2}
$
as the nodal modal intensity. We extract the eigenvectors of the edge frequency dispersions and perform a conjugate product operation on them to obtain the distribution of the eigenstates (see the detailed derivation in Supplementary Material), as shown in Figs.~\ref{fig:3}(a) and~\ref{fig:3}(b), corresponding to nonreciprocal hopping strength $\delta=0.1$ and $\delta=1$ respectively. We also extracted several sets of eigenvectors corresponding to the bulk bands at lower and higher frequencies indicated by blue and red lines respectively, as displayed in Figs.~\ref{fig:3}(c) and~\ref{fig:3}(d). It clearly shows that as the non-Hermitian coefficient increases, the energy of both the edge states and the bulk bands concentrate around the boundary of the electric circuit structure, induced by the non-Hermitian skin effect.

\begin{figure}[!ht]
    \centering
    \includegraphics[width=0.7\linewidth]{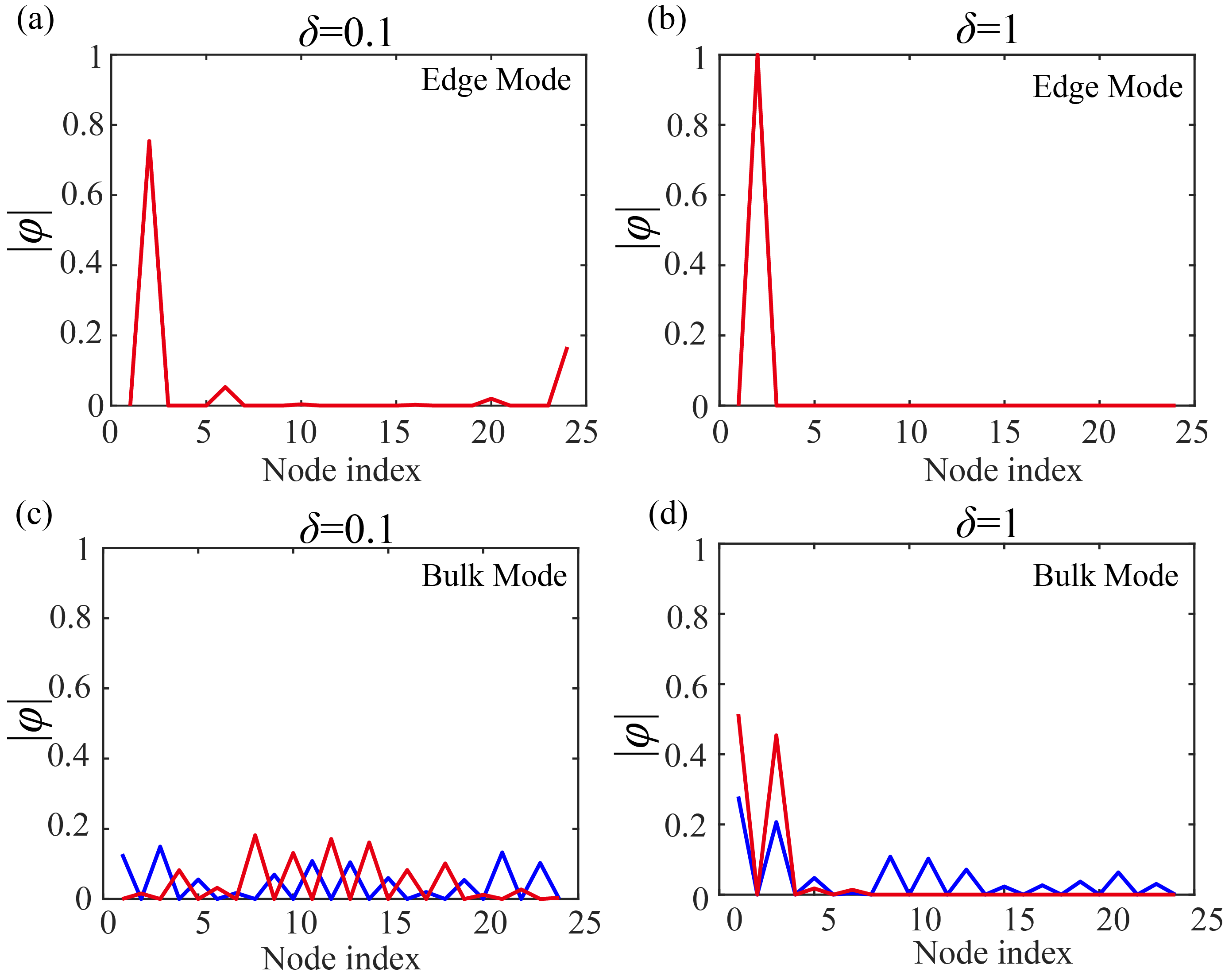}
    \caption{(a) and (b) Eigenstate distribution at the corresponding edge state frequencies in a non-Hermitian topological electric circuit when nonreciprocal hopping strength $\delta=0.1$ and $1$, respectively. (c) and (d) Bulk-state distribution corresponding to bulk band frequencies when $\delta=0.1$ and $1$.}
    \label{fig:3}
\end{figure}

\begin{figure}[!ht]
    \centering
    \includegraphics[width=\linewidth]{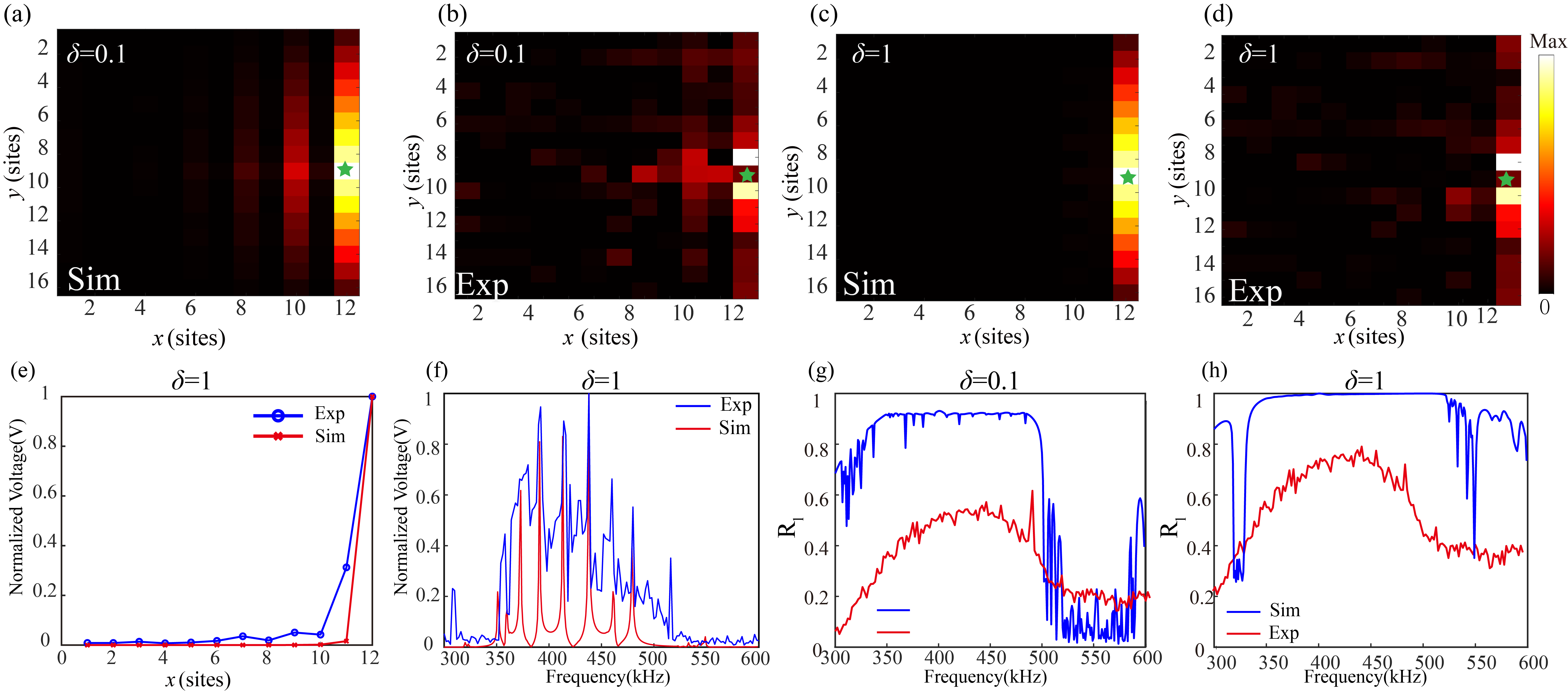}
    \caption{Simulated and experimental distribution of the voltage amplitude within the electric circuit at 486~kHz corresponding to the topological edge state at the nonreciprocal hopping strength $\delta=0.1$ in (a)-(b) and $\delta=1$ in (c)-(d), respectively. The localized edge states are induced by non-Hermitian skin effect. (e) Normalized voltage amplitude distribution along the x direction in (c) and (d). (f) Normalized voltage amplitude as a function of frequencies within 300-600~kHz. (g) and (h) Simulated and measured ratio $R_l$ of edge-localized states.}
    \label{fig:4}
\end{figure}

In order to deeply investigate the properties of the non-Hermitian skin effect in the Möbius topological insulator, we carry out the simulations and experiments in a two-dimensional electric circuit model containing $8\times 6$ unit cells. In the experimental realization, we design and fabricate a printed circuit board (PCB) as shown in Fig.~\ref{fig:1}(b). The excited source is placed in the middle of the right boundary marked by a green star in Fig.~\ref{fig:4}. This configuration aims to enhance the response of the edge states and ensure the effective excitation of the topological modes propagating along the edge. By measuring the voltage response at each node of the electric circuit, we obtain voltage amplitude distributions in the experiments. When the nonreciprocal hopping strength $\delta=0.1$, the simulated and experimental results at 486~kHz show that the voltage amplitude at the right boundary is very large compared with that in the bulk, demonstrating the existence of topological edge states, as displayed in Figs.~\ref{fig:4}(a) and~\ref{fig:4}(b). For the nonreciprocal strength $\delta=1$, the edge state is significantly concentrated in the boundary region of the circuit induced by the non-Hermitian skin effect, as shown in Figs.~\ref{fig:4}(c) and~\ref{fig:4}(d). Figure~\ref{fig:4}(e) demonstrates the edge states exponentially decay perpendicular to the propagating direction for $\delta=1$. The voltage amplitude decays rapidly from the right edge toward the left along x direction, indicating pronounced edge localization. Figure~\ref{fig:4}(f) illustrates the simulated and measured voltage amplitude at one edge node within the frequency range of 300 to 600~kHz corresponding to the twist edge bands. The experimentally observed frequency resonance peaks agree well with simulation results. We do not consider the resistance of the components in the simulated electric circuit, leading to very high resonance peaks. To well make comparison between simulations and experiments, the voltage amplitudes are normalized individually based on their respective maximum values. A minor discrepancy is attributed to the non-ideal characteristics of components such as capacitors, inductors, and circuit wiring, leading to significant resistive losses during the experiment (see the detailed description in Supplementary Material). The operating characteristics of the operational amplifier are highly sensitive to peripheral circuits and are susceptible to a variety of experimental environmental factors such as power supply noise and poor grounding.

To quantify the degree of localization of the edge states, we define~\cite{44}
\begin{equation}
R_{l} = \frac{\int_{\Pi_s} |V|^{2} dx dy}{\int_{\Pi} |V|^{2} dx dy}
\end{equation}

where $|V|$ is the voltage amplitude, $\Pi$ denotes the entire area of the electric circuit, and $\Pi_s$ denotes the area covering the rightmost unit cells in the circuit structure. As illustrated in Figs.~\ref{fig:4}(g) and~\ref{fig:4}(h), the simulated ratio $R_l$ approaches to 1 when the nonreciprocal hopping strength $\delta=1$, demonstrating the strong boundary localization of the edge state within a broad frequency range. The experimentally measured $R_l$ exhibits prominent peaks within the bandgap, and remains low at other frequencies. Both these observations confirm the edge state localization caused by the non-Hermitian skin effect.

\section{Non-Hermitian graphene-like topological semimetals}\label{sec:3}

\begin{figure}[!ht]
    \centering
    \includegraphics[width=0.7\linewidth]{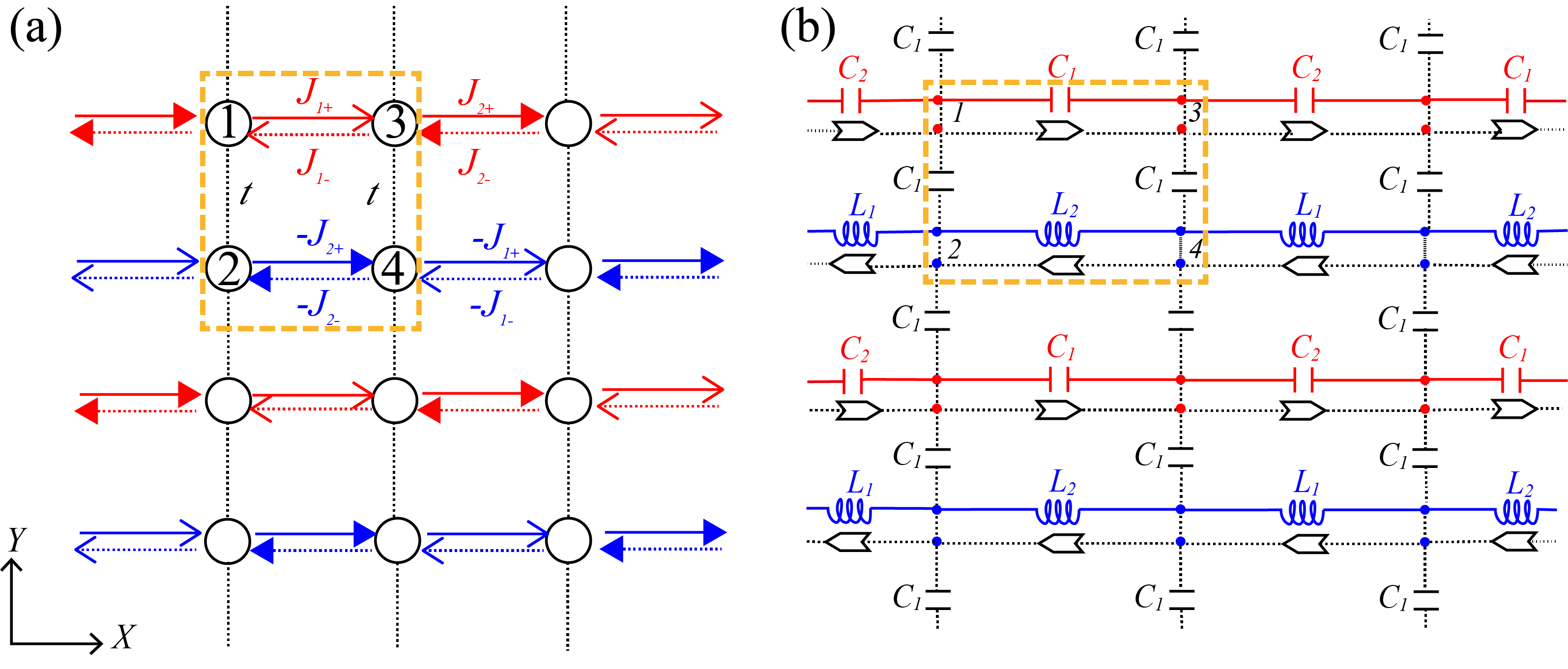}
    \caption{(a) and (b) Tight-binding model and its equivalent circuit model of a non-Hermitian graphene-like topological semimetal.}
    \label{fig:5}
\end{figure}

As shown in Fig.~\ref{fig:5}, we design a rectangular lattice model with alternative dimerization. The primitive translation symmetries along $x$- and $y$- directions are both broken, which splits the fourfold Dirac point into two twofold nodal points~\cite{13}. By introducing the non-Hermitian hopping term based on this rectangular lattice, we construct a non-Hermitian graphene-like semimetal phase with flat bands. The corresponding circuit model is consistent with the construction method of the non-Hermitian Möbius insulator, utilizing capacitors and inductors combined with asymmetric hopping parameters to achieve non-Hermitian characteristics.

The admittance matrix of the non-Hermitian graphene-like semimetal phase is given as follows (detailed derivation in Sec.~I within Supplementary Material):
\begin{equation}
J_{p}=i\omega C_{p}+\frac{1}{i\omega}L_{p}
\label{eq:11}
\end{equation}
where
\begin{equation}
C_{p}=\left[ \begin{matrix}
   t_{2} & A_{y} & B_{x}-\sigma  & 0  \\
   A_{y} & t_{1} & 0 & \sigma   \\
   B_{x}^{*}-\sigma  & 0 & t_{2} & A_{y}  \\
   0 & \sigma  & A_{y} & t_{1}  \\
\end{matrix} \right]
\label{eq:12}
\end{equation}
and
\begin{equation}
L_{p}=\left[ \begin{matrix}
   \frac{1}{L_{3}} & 0 & 0 & 0  \\
   0 & \frac{1}{L_{1}}+\frac{1}{L_{2}} & 0 & -\frac{1}{L_{2}}e^{ik_{x}/2}-\frac{1}{L_{1}}e^{-ik_{x}/2}  \\
   0 & 0 & \frac{1}{L_{3}} & 0  \\
   0 & -\frac{1}{L_{2}}e^{-ik_{x}/2}-\frac{1}{L_{1}}e^{ik_{x}/2} & 0 & \frac{1}{L_{1}}+\frac{1}{L_{2}}  \\
\end{matrix} \right]
\label{eq:13}
\end{equation}

\begin{figure}[!ht]
    \centering
    \includegraphics[width=0.7\linewidth]{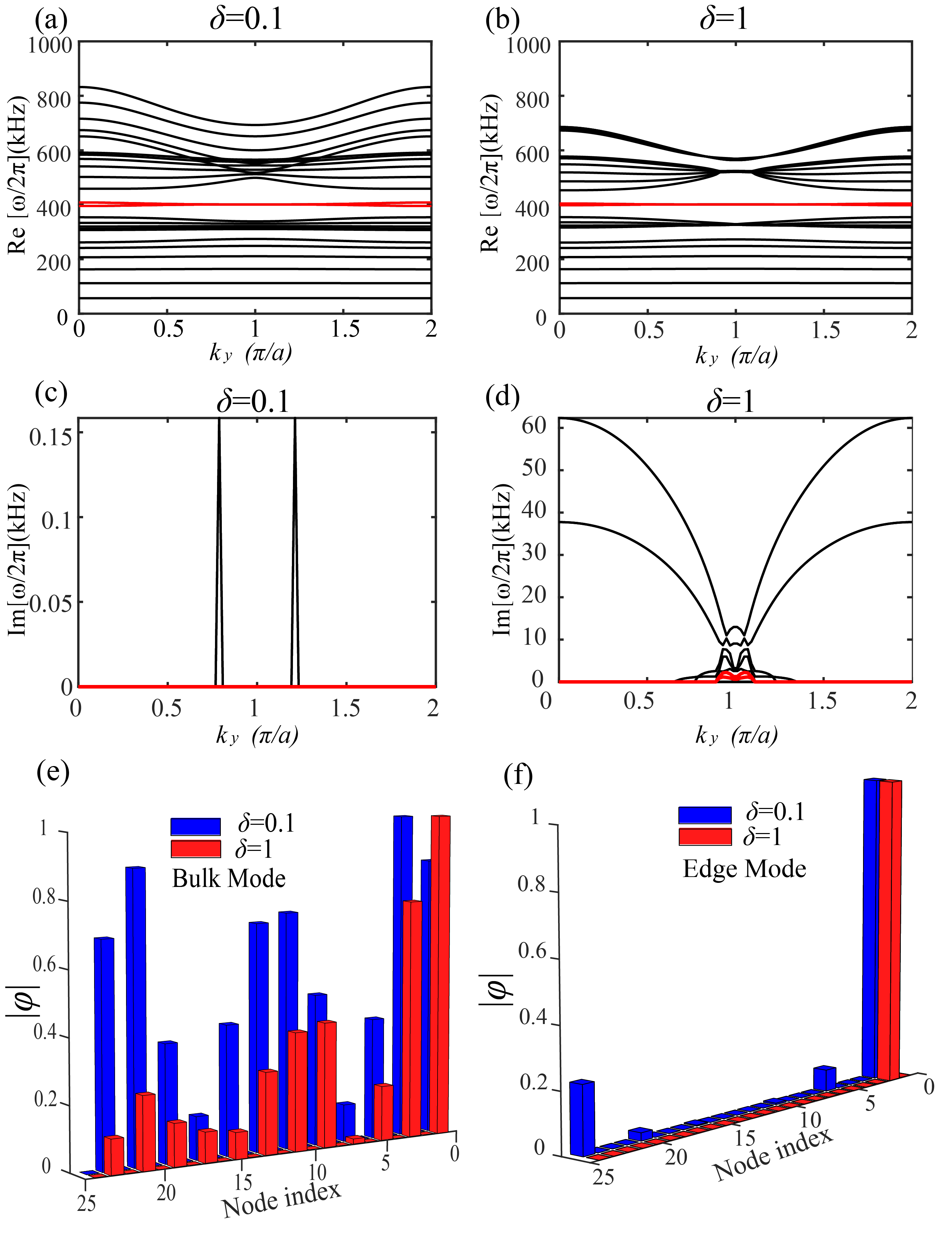}
    \caption{(a) and (b) Dispersion relation of the non-Hermitian graphene-like topological semimetal circuit. Real parts of the frequency are functioned as $k_y$ for the nonreciprocal hopping strength $\delta=0.1$ and $\delta=1$ respectively. (c) and (d) Imaginary part of the frequency, showing enhanced non-Hermitian skin effect as $\delta$ increases. (e) and (f) Normalized energy distribution of bulk and edge states in the graphene-like topological semimetal structure for $\delta=0.1$ and $\delta=1$, respectively.}
    \label{fig:6}
\end{figure}

By solving this eigen-equation, we obtain the band dispersion of the non-Hermitian electric circuit. A relatively flat band, denoted by red lines, emerges within the band gap of the graphene-like topological semimetal, as shown in Figs.~\ref{fig:6}(a) and~\ref{fig:6}(b). We investigate the relationship between the nonreciprocal hopping strength $\delta$ and the frequency spectrum. The result indicates that significant imaginary spectra appear when $\delta$ exceeds 0.15, increasing with larger values of $\delta$, as displayed in Figs.~\ref{fig:6}(c) and~\ref{fig:6}(d). The stronger nonreciprocal hopping term leads to more pronounced non-Hermitian skin effect. Observations of the energy field distribution provide direct evidence for this non-Hermitian skin effect. As depicted in Figs.~\ref{fig:6}(e) and~\ref{fig:6}(f), as $\delta$ increases, the edge state energy becomes highly concentrated at the system boundaries, while the bulk state distribution rapidly decays. This phenomenon aligns with recent discussions on the generalized bulk-boundary correspondence~\cite{23} in non-Hermitian systems, further substantiating the non-Hermitian skin effect in graphene-like topological semimetals.

\begin{figure}[!ht]
    \centering
    \includegraphics[width=0.7\linewidth]{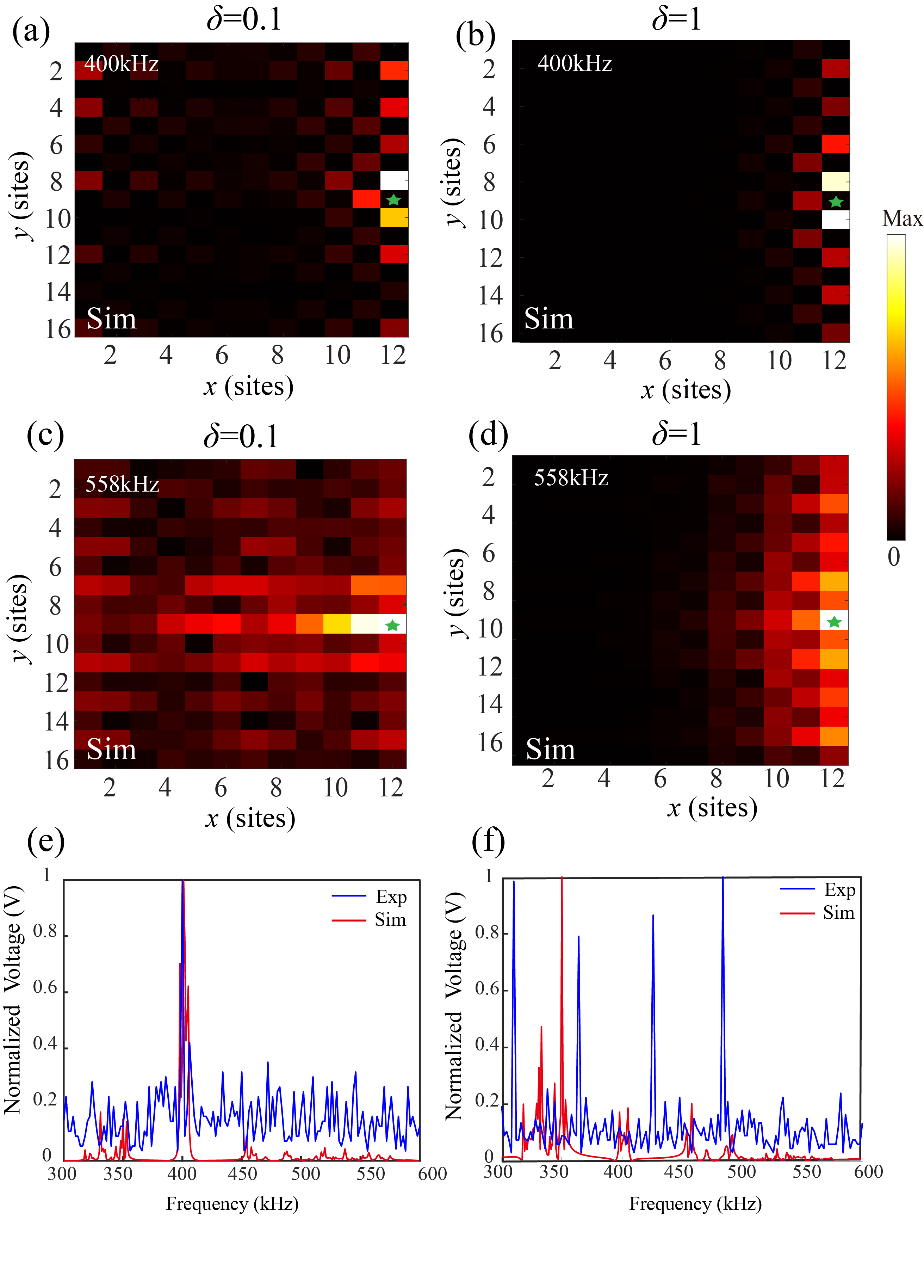}
    \caption{(a) and (b) Voltage amplitude distribution of edge states at 400~kHz in the graphene-like topological semimetal structure for the nonreciprocal hopping strength $\delta=0.1$ and $\delta=1$, respectively. (c) and (d) Voltage amplitude of bulk states at 558~kHz for $\delta=0.1$ and $\delta=1$, respectively. (e) Simulated and experimental normalized voltage amplitude at the edge node as a function of frequency within 300--600~kHz. (f) Simulated and measured normalized voltage amplitude at one bulk node as a function of frequency.}
    \label{fig:7}
\end{figure}

In the experimental design, we employ the same PCB previously utilized in our study of Möbius topological insulators. By flexibly adjusting jumper caps, various topological structures can be achieved without replacing the circuit board itself, thereby enabling precise control over periodic hopping parameters and asymmetric hopping terms. Additionally, the capacitance $C_g$ can be rapidly replaced and adjusted through jumper caps, facilitating a systematic investigation of the impact of non-Hermitian parameters on the topological properties. Simulation results indicate that voltage amplitudes at the boundary are significantly higher than those in the bulk of the electric circuit at 400~kHz, corresponding to the edge state frequency, as shown in Fig.~\ref{fig:7}(a). This clearly demonstrates the significant concentration of topologically protected edge states at the boundary. In contrast, at 558~kHz corresponding to the bulk state frequency, the initial energy distribution spreads into the bulk, as shown in Fig.~\ref{fig:7}(c). However, with the introduction of non-Hermitian hopping terms, the energy rapidly redistributes towards the edges, exhibiting a pronounced non-Hermitian skin effect, as illustrated in Figs.~\ref{fig:7}(b) and~\ref{fig:7}(d). Notably, this skin effect is not limited to edge states but can be observed in bulk states. Figure~\ref{fig:7}(e) illustrates the simulated and experimentally measured normalized voltage amplitudes at one edge node as a function of frequencies. The highest voltage peak corresponds to the frequency of the flat edge state dispersion. Figure~\ref{fig:7}(f) displays the irregular normalized voltage peaks at one bulk node. With further increases in the non-Hermitian parameter, the energy tends to accumulate more at the edges for both edge and bulk states. This phenomenon indicates that the non-Hermitian hopping term not only enhances the bulk-boundary correspondence of the system but also profoundly reshapes the spatial distribution of the overall energy field. These findings strongly confirm the universality of the non-Hermitian skin effect and its significant impact on the dynamical behavior of the system.

To better understand the non-Hermitian skin effect in our circuit models, we compare its mechanism and manifestation with those of representative non-Hermitian systems, including the Hatano-Nelson model~\cite{45,46,47} and the non-Hermitian Su-Schrieffer-Heeger (SSH) model with nonreciprocal hoppings~\cite{48,49}. In these models, the non-Hermitian skin effect arises from asymmetric hopping amplitudes, leading to unidirectional mode localization and spectral winding in the complex plane. Our non-Hermitian Möbius circuit exhibits the non-Hermitian skin effect as a result of broken translation symmetry, allowing us to study how twisted edge dispersions are deformed under nonreciprocal couplings. Meanwhile, in the non-Hermitian graphene-like semimetal, bulk states and edge states develop edge localization, especially near the flat-band region, revealing a bulk-band skin effect.

\section{Conclusion}\label{sec:4}

This study provides a systematic and in-depth investigation into the realization and characteristics of non-Hermitian Möbius insulators and graphene-like topological semimetals implemented within circuit-based platforms, incorporating theoretical modeling, numerical simulations and experimental verification. The results conclusively demonstrate that introducing nonreciprocal hopping mechanisms is a crucial condition for achieving non-Hermitian topological states. Moreover, the tuning of non-Hermitian parameters not only markedly enhances the energy localization of topological edge states and the intensity of the non-Hermitian skin effect, but also highlights the universality of such topological features within rectangular lattice configurations. These findings provide robust theoretical and experimental support for the design and optimization of non-Hermitian topological materials. More importantly, this work bridges the gap between Hamiltonian models of non-Hermitian topological phases and experimentally realizable circuit-based systems. Such an approach enables precise experimental validation of theoretical predictions and lays a solid foundation for extending these concepts to a broader array of applications, including photonic/acoustic topological devices and quantum simulation platforms. The electric circuit framework and analytical methods develop herein not only serve as a general paradigm for future systematic research and engineering optimization of non-Hermitian topological systems, but also chart a new course for the advancement of topological circuits and related emerging fields. Our study provides a platform to investigate the interplay between non-Hermiticity, topological band engineering, and boundary localization. Beyond its fundamental interest, the observed non-Hermitian skin effect and its controllability in electric circuits may also have practical implications. Given the recent successes of non-Hermitian circuit-based sensors with ultrahigh sensitivity~\cite{50,51,52}, we believe that incorporating non-Hermitian skin effect and edge state dispersions of both Möbius-type and flat-band-type could offer new opportunities for designing robust and tunable sensing devices on both PCB and chip platforms.

\textbf{Note added.} -- After submission, we became aware of Ref.~\cite{53}, realizing the non-Hermitian acoustic Möbius insulator by introducing different loss materials, which differs for our proposed Non-Hermitian Möbius topological circuit work by introducing the nonreciprocal hopping term.

\bibliographystyle{unsrt}  
\bibliography{manuscript}

\end{document}